\documentclass[aip,jcp,preprint,twocolumn]{revtex4-1}

\usepackage{amsmath}
\usepackage{amsfonts}
\usepackage{amssymb}
\usepackage{graphicx}

\begin{document}

\title{
Development of a fragment kinetic Monte Carlo method 
for efficient prediction of ionic diffusion in perovskite crystals 
}

\author{Hiroya Nakata}
\email{hiroya.nakata.gt@kyocera.jp}
\affiliation{Kyocera Corporation, Research Institute for Advanced Materials and Devices, 3-5-3 Hikaridai Seika-cho Soraku-gun Kyoto 619-0237, Japan.}

\begin{abstract}
   A massively parallel kinetic Monte Carlo (kMC) approach is proposed
   for simulating ionic migration in a crystal system by introducing 
   the atomic fragmentation scheme (fragment kMC). 
   The fragment kMC method achieved a reasonable 
   parallel efficiency with 1728 central processing unit (CPU) cores, 
   and the method enables the simulation of ionic 
   diffusion in $\mu$m-scale perovskite crystals.  
   To demonstrate the feasibility of the proposed approach,
   the fragment kMC method was applied to predict the diffusion coefficients 
   of hydride ions and oxygen ions in SrTiO$_{(3-x)}$H$_x$ 
   and BaTiO$_{(3-x)}$H$_x$ systems.
   Finally, the fragment kMC method was customized for 
   $\mu$-scale BaTiO$_3$ simulation under an applied bias voltage,
   and oxygen diffusion in BaTiO$_3$ model  was 
   evaluated. The respective grain sizes are sub-nanometre,
   and we conclude that the proposed fragment kMC method can 
   be applied to calculate the extent of ionic migration
   in  $\mu$-scale materials with fully atomistic simulation 
   models at a reasonable computational cost.
\end{abstract}


\maketitle



\section{Introduction}
  Perovskite material, which is typically composed of an ABO$_3$ 
  type stoichiometry with a cubic crystal structure, draws considerable attention,
  because of its applicability in many electronic devices that are important for industry.
  The potential application fields include
  sensors\cite{Sensor00,SensorSTO}, 
  random access memory\cite{nagaraj1999ba,yoo2015resistive,STOmemory},
  batteries\cite{zhao2012hierarchical,suntivich2011design}, 
  capacitors\cite{kishi2003base,Randall00},
  piezoelectric devices\cite{Randall01,park2014highly},
  solid fuel cells\cite{ishihara2009perovskite},
  and catalysts for the water-splitting reaction\cite{yang2014thermodynamic}.

  In all these applications, the stability and migration of oxygen vacancies 
  are a key factor for determining the utility of the electronic devices.
  For example, the blocking the oxygen vacancies diffusion by the grain boundaries 
  is one of the important factor for the long-term failure of multilayer ceramic 
  capacitors (MLCCs), and a number of experimental studies 
  of the degradation mechanism have been reported\cite{
        yang2004oxygen,chazono2001dc,chazono2003effect,hennings2001dielectric, 
        albertsen1998donor,strukov2012thermophoresis,yang2003modulated, 
        waser1989electrochemical}.
  For these reasons, experimental investigations of oxygen diffusion 
  in perovskite oxides played a central role in designing highly functional 
  materials for industrial devices\cite{shirasaki1980defect,itoh2002effect,itoh2002oxygen, 
                 hasegawa1991enhanced,fromling2011oxygen,kessel2010strongly,
                 de2015oxygen}.

  Until recently, ionic migration was empirically predicted
  according to the ionic radius and size of the unit cell.
  With the development of computer science,
  a more detailed analysis of the diffusion mechanisms can be performed 
  for a more sophisticated design towards the development of highly functional materials.  
  Therefore, theoretical simulations are becoming powerful tools for understanding 
  the vacancy-mediated ionic diffusion. 
  In particular, quantum mechanics (QM)-based analysis of vacancy 
  diffusion in perovskite  crystal structures has been performed intensively
  for BaTiO$_3$\cite{BaTiO3QM1,BaTiO3QM2,BaTiO3QM3,munch1997quantum},
      SrTiO$_3$\cite{zhang2016oxygen,STOOdiffusion},
      BaZrO$_3$\cite{shi2005first,bjorketun2007effect},
      and many other perovskite crystal structure materials\cite{kreuer2003proton,kreuer1999aspects,kang2013first}.
  The above QM approaches were mostly limited to static analysis of phenomena such as 
  formation enthalpy and activation barriers because of the high computational 
  cost. To focus on the kinetics of the ionic migration, the stability of vacancies 
  near the grain boundary, crystal dislocation, or crystal transition, 
  a number of molecular dynamics (MD) simulations were performed 
  for the bulk\cite{zhang2018atomistic,zhang2019optimising,tealdi2010layered,
                saifulaislam2008atomic,parfitt2011oxygen},
       surface or interface\cite{waldow2016computational}, 
       and dislocations\cite{STOMD}.
   Although the above MD simulations can cover most ionic migration behaviours,
   the applications are limited within the ionic diffusion of oxygen in the relatively 
   high-temperature region (more than 1000~K), the system size is within
   several nm$^3$, and the simulation time scale is several ns.
   However, for a typical electronic device operating environment,  
   the temperature range is between 300~K and 500~K,
   and the material sizes are around 1~$\mu$m.
   Thus, it is necessary to develop an alternative approach  
   to simulate the ionic migration in such low-temperature regions     
   with much larger computational models.

   The kinetic Monte Carlo (kMC) 
   method\cite{kMCbook,kMCbasis01,kMCbasis02} is 
   one of the promising 
   approaches towards such large-scale ionic migration simulation.
   kMC enables one to calculate long-time scale simulations, 
   in which the vacancy jumps into a different nearest site
   based on the predefined transition probability.  
   kMC has been applied to predict the kinetics of many problems, 
   such  as yttria-stabilized zirconia\cite{lau2009kinetic,lau2008kinetic,tada2013},
   oxygen diffusion in doped ceria\cite{ceria01,ceria02},
   chemical reactions\cite{kMCCO01,kMCN01,kMCwat01,KMCwat02,kMCNH301,kMCNH302,kMCMeOH},
   material diffusion\cite{kmcOld01,matera2011adlayer,temel2007does,rieger2008effect,kMCSOFC},
   electrochemical impedance\cite{pornp2007electrochemical},
   chemical catalysis\cite{hansen1999modeling,hansen2000first,hansen2000first2,reuter2006first,stamatakis2011first,boscoboinik2008monte,kunz2015kinetic},
   adsorbate--adsorbate interactions,\cite{liu2013realistic,stamatakis2016rationalizing},
   and
   crystal growth\cite{gilmer1980computer,schulze2004hybrid}.
   Despite the success of the kMC approach, there are basically 
   two limitations to applying the method to oxygen migration in
   perovskite crystal structures. First, the simulation time scale decreases
   with the increase in system size. 
   Second, the efficient parallel programming of ionic migration 
   for kMC is difficult. The update of the event list and their selection 
   can be straightforwardly estimated at minor computational cost,
   but the bottleneck is the communication between the boundary regions 
   of the respective parallelized lattice. 
   Thus, despite the mentioned success of the kMC approach, 
   a more sophisticated approach is necessary towards the simulation of 
   ionic migration in $\mu$m-scale ceramic materials.
   
   To reduce the computational cost of kMC approach,
   several group developed parallel version of kMC 
   based on the block separation,
   which separate the kMC event selection based on the respective 
   predefined block groups\cite{li2019openkmc,tada2013}. 
   Even these  success of the parallel approach, 
   the method require further development to simulate 
   much more complex chemical reaction and ionic diffusions. 
   In this study, we propose a highly efficient approach of 
   ionic migration simulation, in which the kMC simulation is fragmented 
   based on the respective atoms (fragment kMC). 
   Then, the complex event selection
   and its parallelization can be decomposed into the respective 
   fragments. The fragment kMC approach shows a 
   reasonable parallel efficiency with a large number of central processing unit (CPU) cores. 
   Subsequently, we discuss the effectiveness of the fragment kMC approach
   by applying the method to predict the diffusion of the hydride ion
   in the SrTiO$_{(3-x)}$H$_x$ and BaTiO$_{(3-x)}$H$_x$ systems.
   It is experimentally known that diffusion of the hydride ion is 
   vacancy-mediated,
   and its diffusion coefficient depends significantly 
   on the concentration of hydride ions\cite{liu2019highly}. 
   We demonstrate that the fragment kMC approach can straightforwardly 
   predict the diffusion coefficients in such complex ionic migrations. 
   Second, the oxygen migration in BaTiO$_3$ was evaluated under applied 
   voltage. To make a $\mu$-scale  model, 
   the computational model contained 7,680,000,000 atoms, 
   and the simulation time was 5~ms. 
   In this study, the effectiveness of fragment kMC was demonstrated 
   by performing vacancy migration for several specific examples 
   related to energy and electronic devices, and we opened up a new 
   territory of application of atomistic material simulation 
   in large-scale diffusion analysis.

\section{Theoretical method}
\subsection{Summary of the conventional  kMC method}
  In this section, we briefly summarize the conventional 
  kMC approach, and more details can be found in 
  many places\cite{kMCbook,kMCbasis01,kMCbasis02}.
  In the jump-diffusion kMC approach, 
  the transition rate for the $i$th event, $k_i$,  is evaluated 
  based on the rate constant of the chemical reaction as follows:
  \begin{align}
    k_i = A 
        \mathrm{exp}
        \left[
          - \frac{E_a}{RT}
        \right],
    \label{RateCnst1}
  \end{align}
  where $E_a$, $R$, $T$, and $A$ are the activation energy, universal gas constant,
  temperature, and pre-exponential factor, respectively.
  Then, the transition probability for the  $i$th event $p_i$ is estimated by
  \begin{align}
    p_i = \frac{k_i}{\sum_j k_j}.
    \label{TransitionProb}
  \end{align} 
  Based on the probability $p_i$, one of the events is chosen, 
  and the geometry is updated according to the selected event.
  To select an event, typically, 
  a random number $r$ is generated, 
  and then one can define the selected event $l$ 
  so as to satisfy the following equation:
  \begin{align}
     \sum_i^l k_i  / \sum_i^N k_i
     \le r 
     \le 
     \sum_i^{l+1} k_i / \sum_i^N k_i,
     \label{eventSelection}
  \end{align}
  where $N$ is the number of total event.
  The geometry and the transition probability $p_i$ for respective events are  
  updated according to the selected event $l$,
  and $n+1$th step simulation time $t_{n+1}$ is also updated 
  by another random number, $r^\prime$,
  as follows:
  \begin{align}
     t_{n+1}  = t_n - \frac{ln(r^\prime)}{\sum_i k_i}.
     \label{TimeStep}
  \end{align}
  
  Despite its simple theory and algorithm, the application of kMC to a large 
  system is sometimes difficult, because there are basically two bottlenecks
  for practical application of the kMC simulation.
  First, it is not difficult to see that the time step shown in Eq.~(\ref{TimeStep}) 
  becomes smaller and smaller when the number of events in the system increases. 
  Second, the sequential event selection and event update
  are difficult to parallelize. As a result, the application territory 
  of the standard kMC method is limited within a relatively small size of 
  the computational model and a short simulation time in the case of 
  the jump-diffusion approach.

\subsection{Atomic-based fragment kMC method}
  Towards the simulation of microscale kinetics of atomic diffusion and 
  its massive parallel algorithm, 
  the rate constant of the chemical 
  reaction in Eq.~(\ref{RateCnst1}) can be rewritten with atomic-based formulation 
  as follows:
  \begin{align}
    k_i^I = A 
        \mathrm{exp}
        \left[
          - \frac{E_a}{RT}
        \right],
  \end{align}
  where $I$ is the atomic index.
  Then, the total rate for each atom $R_I$ and the maximum rate 
  constant $R^\mathrm{max}$ are 
  \begin{align}
    R_I             =&  \sum_{i \in I} k_i^I,
  \\
    R^\mathrm{max}  =&  \mathrm{max} 
                        \left(
                          R_I
                        \right).
  \end{align}
  Using the maximum rate constant $R^\mathrm{max}$, 
  the transition probability $p_i$ is reformulated as  
  \begin{align}
    p_i = \frac{R^\mathrm{max}}{\sum_J \sum_j k_j^J} 
          \frac{k_i}{R^\mathrm{max}},
  \end{align} 
  and the probability to select atom~$I$ ($p^I$) is
  \begin{align}
    p^I = &
        \sum_{j \in I} p_j^I
        \\
        = &
        \frac{R^\mathrm{max}}{\sum_J \sum_j k_j^J} 
        \sum_{j \in I} 
        \frac{k_j}{R^\mathrm{max}}
        \label{AtomP}.
  \end{align} 
  The first term in Eq.~(\ref{AtomP}) is independent of fragment $I$,
  which means that the selection probability only depends on 
  the rate constants related to atom~$I$.
  This concept of dynamic renormalizatin can be also found in 
  the study Grieshammer et. al.\cite{kMCRenormalize}.

\begin{figure}
          \begin{center}
            \includegraphics[clip,width=12.0cm]{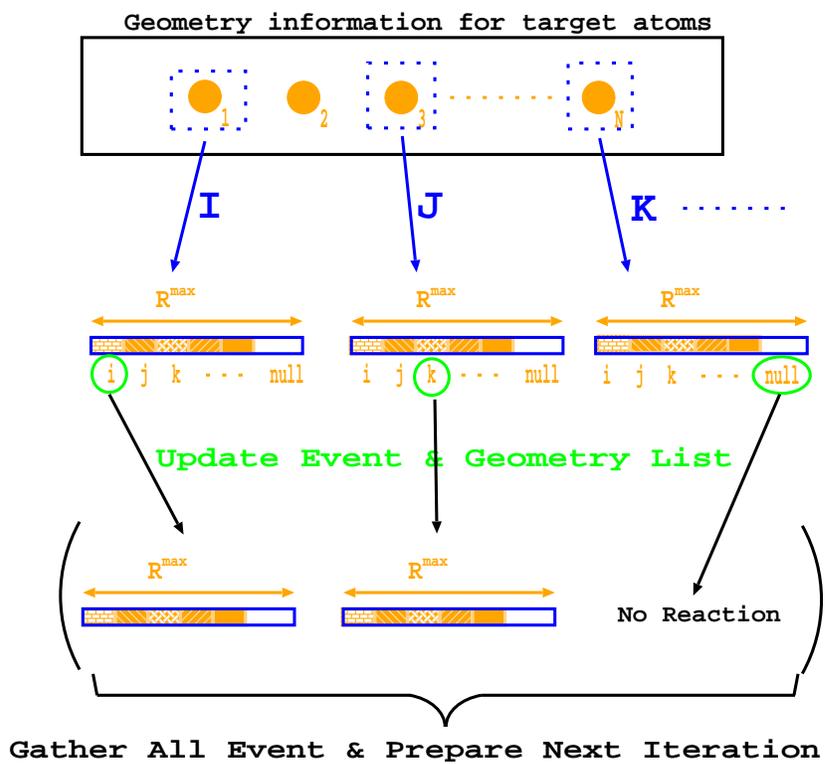} \\
          \end{center}
	      \caption{
              Schematic illustration of how the fragment kMC method updates 
              the event selection and the geometry update process.
              \label{Figure01}
		      }
	\end{figure}

  By introducing the auxiliary value $R^\mathrm{max}$, 
  the event selection can be decomposed 
  into selection of atoms and of events. 
  Then, the kMC algorithm can be reformulated
  (see Figure~\ref{Figure01}).
  First, $N^\mathrm{frg}$ atoms in the system are selected
  randomly (there are three selected atoms in Figure~\ref{Figure01} 
  for example). 
  Then, for each selected atom, the events are chosen based 
  on the transition probability $\frac{k_i}{R^\mathrm{max}}$.
  If the total transition rate constant $\displaystyle{\sum_{i\in I} k_i}$ 
  is less than $R^\mathrm{max}$, 
  we introduce a vacancy event region, 
  where no atomic transition and no time integration
  occur.
  With the pseudo-event selection, 
  the probability of selecting each event is exactly the same as  
  the original formulation of the transition probability 
  shown in Eq.~(\ref{TransitionProb}), and each event can be 
  updated independently.

  This fragmentation of the event update procedure 
  enables performing an efficient parallel implementation of
  kMC for atomic diffusion or for a chemical reaction. 
  Then, all the related event lists are gathered, and 
  the rate constant for the next stage of event 
  selection is updated.
  Simultaneously, the next step simulation time $t_{n+1}$ is incremented 
  as follows:
  \begin{align}
     t_{n+1}  = t_n - \frac{ln(r^\prime)}{\sum_i k_i} N^\mathrm{updated},
     \label{TimeStepFrg}
  \end{align}
  where $N^\mathrm{updated}$ is the event number
  except that the selected event is null.
  Thus, if the number of the updated event $N^\mathrm{updated}$ is 
  more than one, the event and corresponding atoms are updated 
  independently.

\section{Computational details}
  The fragment kMC approach was implemented into the kMC programme 
  (written in C++), and the programme was parallelized with 
  message passing interface (MPI).
  (the fragment kMC programme is available free of charge
  at GitHub (https://github.com/hiroyanakata/kMC.v01)).

\begin{figure}
          \begin{center}
           \includegraphics[clip,width=12.0cm]{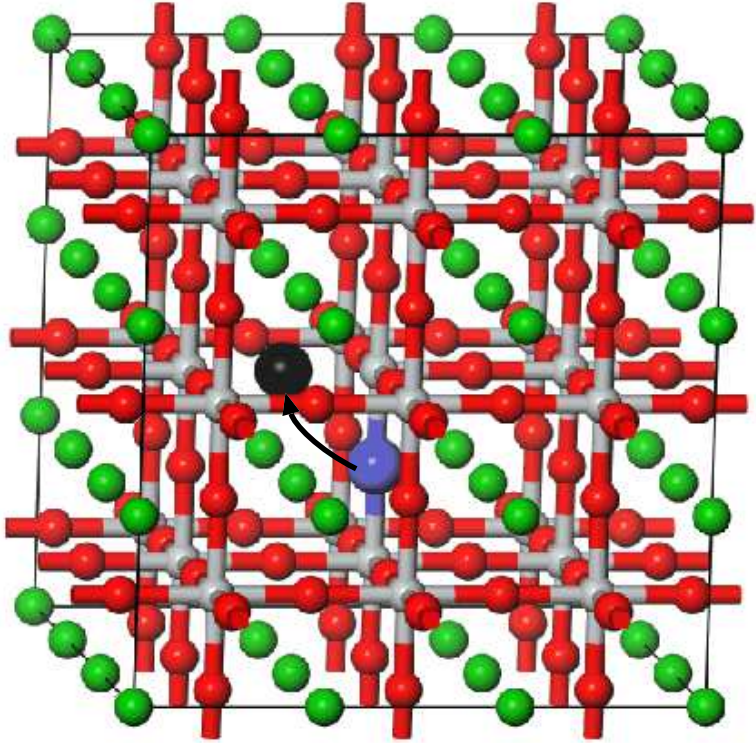} \\
          \end{center}
	      \caption{
              Crystal structure of the Ba(Sr)TiO$_3$ system 
              used in this study.
              Green, red, and gray denote the Ba(Sr), O, and Ti, respectively.
              The black denote the oxygen vacancy, and the blue oxygen
              migrate to the black vacancy site during the kMC simulation.
              \label{Figure02}
		      }
	\end{figure}

  As for the pilot test of the developed programme, 
  the vacancy diffusion in SrTiO$_3$ and BaTiO$_3$ single perovskite 
  crystals was evaluated (see Figure~\ref{Figure02} for the crystal structure).    
  The hydride ions diffusion in SrTiO$_3$ was recently reported 
  by Liu et al.\cite{liu2019highly}, but the diffusion coefficient 
  for each hydride and oxygen ion separately is not yet clearly understood. 
  Thus, in this study, the hydride ions diffusion in perovskite 
  SrTiO$_{(3-x)}$H$_x$ and BaTiO$_{(3-x)}$H$_x$ was also evaluated,
  where $x=$ 0.25, 0.35, and 0.45.
  The activation energies $E_a$ for the vacancy diffusion 
  of oxygen are 0.6~eV for SrTiO$_3$\cite{de2012behavior,de2015oxygen}
            and 0.7~eV for BaTiO$_3$\cite{BaTiO3expt},
  which were verified by experiment.
  Because there is no experimental study of the activation barrier 
  of a pure hydride ions, the activation energy of hydride ions
  was estimated from a previous first principle calculation
  \cite{liu2019highly,liu2017formation}; 
  the activation barrier of hydride ions was chosen as 0.17 and 0.28~eV for SrTiO$_3$ 
  and BaTiO$_3$, respectively.  
  For all the simulation, the pre-exponential factor $A$ is set 
  to 1.0e13 (1/s).

  First, the accuracy of the developed fragment kMC 
  was evaluated by comparing with the standard kMC. 
  Then, the parallel efficiency of fragment kMC was evaluated
  by performing a million-step kMC simulation,
  and the computational timing was evaluated
  from a single core to 1728~cores. 
  The system size was 600$^3$~unit cells for the performance tests, 
  while we used a relatively large system size (1200$^3$~unit cells)
  for evaluation of the parallel efficiency from 32 to 1728 cores; 
   0.1\% of oxygen atoms in the crystal were replaced with vacancies.
 
  Second, the hydride ions and oxygen diffusion coefficients were
  evaluated for the temperature range from 550 to 700 K. 
  Then, the apparent activation barrier was evaluated 
  by the Arrhenius plot. 
  In this study, the diffusion coefficients for 
  hydride ions and oxygen were separately evaluated by
  tracking the trajectory of the respective atoms.

  Finally, to demonstrate the effectiveness of the fragment kMC approach,
  the oxygen vacancy diffusion in $\mu$ scale BaTiO$_3$ was evaluated 
  under an applied voltage.
  To consider the effect of the applied voltage,
  the electrostatic potential was considered by  solving the Poisson equation,
  and the effect of electrostatic potential (ESP) on the rate constant
  was included as follows:
  \begin{align}
    k_i^I = A 
        \mathrm{exp}
        \left[
          - \frac{E_a + \Delta E_\mathrm{pot}}{RT}
        \right],
  \end{align}
  where $\Delta E_\mathrm{pot}$ is the ESP difference 
  between product and reactant position. 
  The system size was 641.6~nm $\times$ 160.4~nm
  $\times$ 962.4~nm, which is nearly the experimental 
  size of a ferroelectric material, and the total number 
  of atoms was 7,680,000,000.
  To run such a large-scale computational model,
  the simulation was performed with 1536 CPU cores. 
  The detailed calculation model is shown 
  in the next section.

\section{Results and discussion}

\subsection{Accuracy and parallel efficiency of the fragment kMC method}
  To investigate the accuracy of the fragment kMC approach,
  the simulation results of vacancy diffusion were compared 
  between different numbers of fragments from 
  $N^\mathrm{frg}=1$ (conventional kMC) to  $N^\mathrm{frg}=216$,
  and the parallel efficiencies of the respective fragment patterns 
  were evaluated to demonstrate the effectiveness of 
  the developed approach.
  For this purpose, the vacancy diffusion coefficients in SrTiO$_3$
  and SrTiO$_{2.75}$H$_{0.25}$ were investigated.   

\begin{figure}
          \begin{center}
           \includegraphics[clip,width=8.0cm]{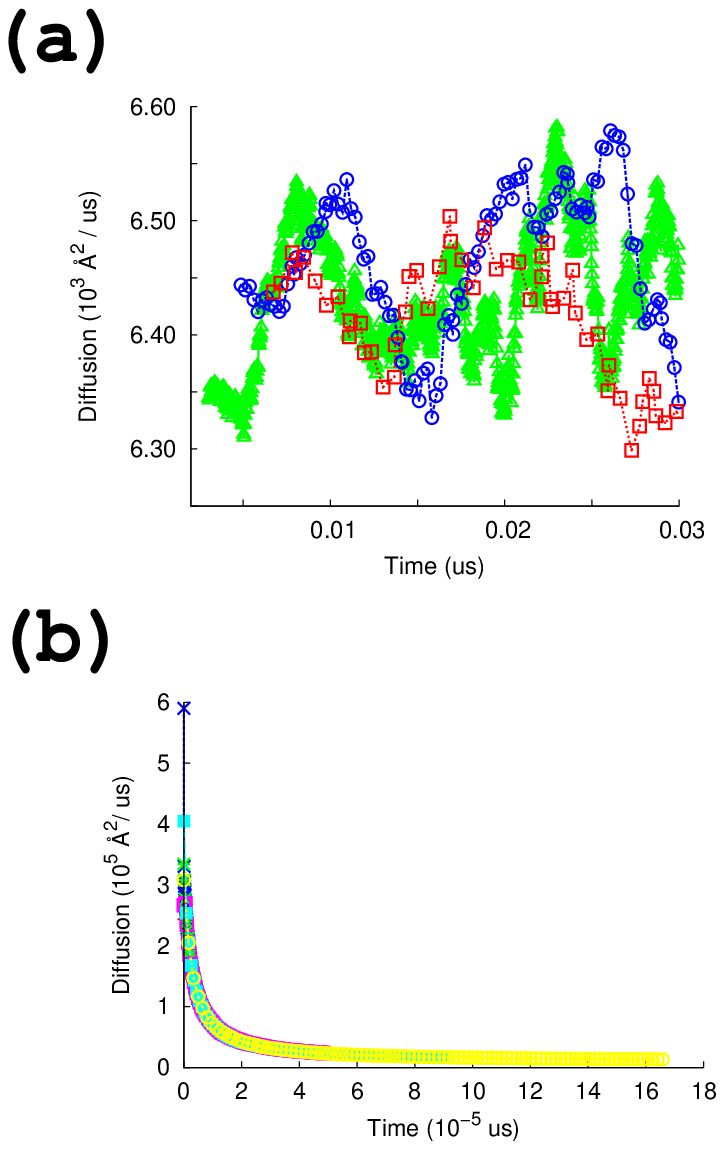} \\
          \end{center}
	      \caption{
               Comparison between the vacancy diffusion coefficients 
              obtained with different numbers of fragments.
              The unit of the vacancy diffusion coefficient is \AA$^2$/us.
              (a) SrTiO$_3$: The red, blue, and green lines denote
                  the results with 216 and 80 fragments and conventional kMC, respectively.
              (b) SrTiO$_{(3-x)}$H$_x$  with $x=0.25$:
                  Red, blue, green, magenta, sky blue, and yellow denote
                  the results with conventional kMC and 1, 8, 32,
                  64, 125, and 216 fragments, respectively. 
              \label{Figure03}
		      }
	\end{figure}

  The simulation results for various numbers of 
  fragments $N^\mathrm{frg}$ are summarized in Figure~\ref{Figure03}(a) and (b)
  for SrTiO$_3$
  and SrTiO$_{2.75}$H$_{0.25}$, respectively.
  In Figure~\ref{Figure03}, the different colours denote 
  the vacancy diffusion coefficient of the respective fragment number. 
  
  As shown in Figure~\ref{Figure03}(a),
  the diffusion coefficients in SrTiO$_3$ agree well with each other.
  The average and standard deviations of the oxygen vacancy diffusion coefficient
  are shown in Table~\ref{TABLE01}. The differences between the respective 
  fragmentation types are only within several nanometres/microseconds, which is 
  less than the value of the standard deviation.

  In the SrTiO$_{2.75}$H$_{0.25}$ system, the hydride and oxygen ions 
  have two different transition barriers, with activation barriers
  of 0.17 and 0.6~eV, respectively. 
  The event selection of hydride and oxygen ions can be understood
  based on the correlation factor of the hydride ion\cite{liu2019highly}.   
  One can estimate the reaction rate for each hydride or oxygen ion 
  based on eq.~\ref{RateCnst1},
  and the transition rates of hydride ions are about 3000 times larger than 
  those of oxygen ions. 
  Thus, the vacancy transition was dominated by hydride ions,
  and the initial large diffusion coefficient of vacancy 
  is attributed to the diffusion coefficient of hydride ion itself ($D_H$).
  On the other hand, the diffusion of hydride ion in equilibrium state ($D_{H}^*$)
  is highly correlated with the migration of oxygen ion as follows:
  \begin{align}
     D_{H}^* =& f_H D_H
  \\
     f_H     =& \frac{\frac{2}{1-f}(\frac{r_H}{r_O}\chi_H f_H + \chi_O f_O)}
                    {2\frac{r_H}{r_O} 
                    + \frac{2}{1-f}(\frac{r_H}{r_O}\chi_H f_H + \chi_O f_O)},
     \label{Cfactor}
  \end{align}
  where $f_H$ and $f_O$   are  the correlation factor for hydride and oxygen ion,
  and  $\chi_H$, and $\chi_O$ are  fractional concentration of hydride and oxygen ion.
  The correlation factor  can be estimated by first principle simulation
  (See ref.~\cite{liu2019highly,AtomMovementTheory} for more detail). 
  In the target temperature, the value of $f_H$ is order $10^{-3}$.
  Thus, the actual hydride ion diffusion coefficients ($D_{H}^*$) is significantly 
  less than the diffusion coefficients of hydride ion itself ($D_H$),
  and we observed such a decrease of diffusion coefficients of 
  hydride ion in Figure~\ref{Figure03}(b).
  In comparison with conventional kMC approach,
  such vacancy diffusion with a mixture of different transition
  barriers can also be reasonably reproduced by the fragment kMC approach.
  
\begin{figure}
          \begin{center}
           \includegraphics[clip,width=12.0cm]{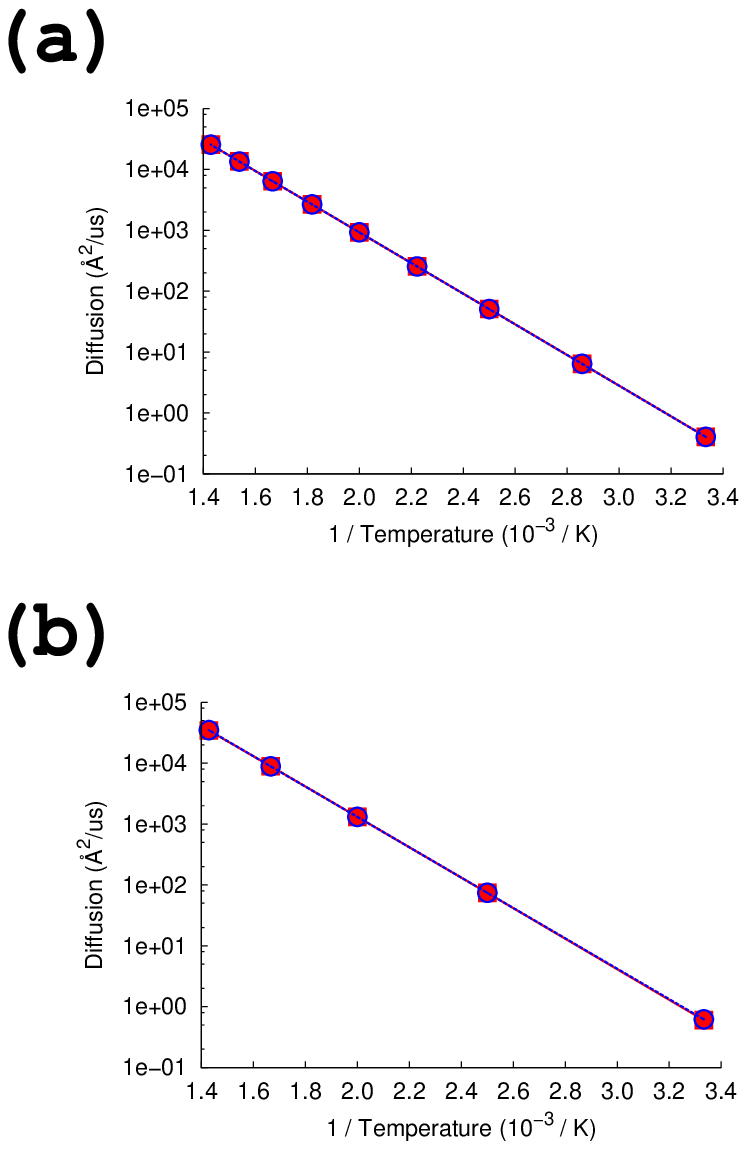} \\
          \end{center}
	      \caption{
              The average vacancy diffusion coefficient 
              for different temperatures. Comparison 
              between (blue) standard kMC and (red) fragment kMC 
              for (a) SrTiO$_3$ and (b) SrTiO$_{(3-x)}$H$_x$  with $x=0.25$. 
              The number of fragments is 80.
              \label{Figure04}
		      }
	\end{figure}

  Furthermore, the temperature dependence of the vacancy diffusion
  coefficient was evaluated to demonstrate the accuracy of the fragment 
  kMC method. For this purpose, the vacancy diffusion coefficients in both
  SrTiO$_3$ and SrTiO$_{2.75}$H$_{0.25}$ were evaluated for the temperatures
  500, 550, 600, 650, and 700~K, respectively.
  A comparison of the results between the conventional kMC
  and fragment kMC (where the number of fragments is 80) approaches
  is shown in Figure~\ref{Figure04}.
  A perfect agreement is observed
  between the results with and without fragmentation for both SrTiO$_3$ and 
  SrTiO$_{2.75}$H$_{0.25}$. The agreement between the diffusion coefficients 
  with and without the fragmentation approach  
  also proves that the developed fragment kMC method
  works for simulation of vacancy diffusion of oxygen and hydride ions.
   
\begin{figure}
          \begin{center}
           \includegraphics[clip,width=12.0cm]{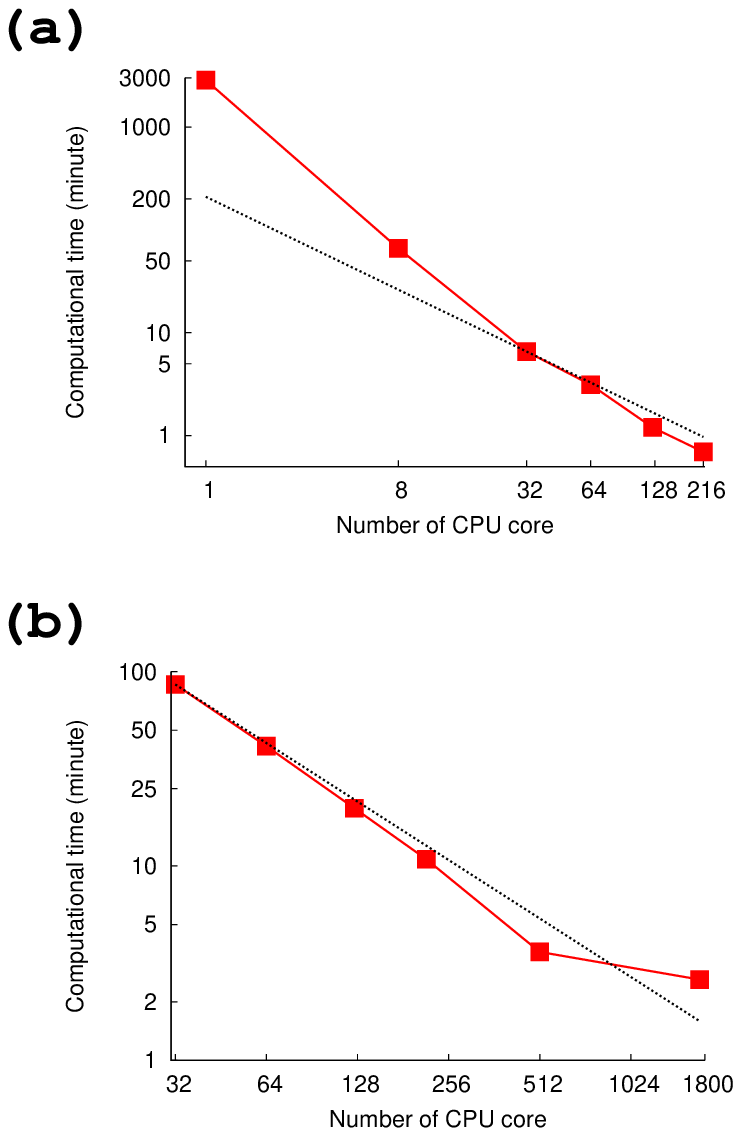} \\
          \end{center}
	      \caption{
              Computational timing and its parallel efficiency of fragment kMC.
              The black dashed line is the ideal computational time
              estimated by using the computational time with 32 cores. 
              (a) for 600 $\times$ 600 $\times$ 600 unit cells from 1 to 216 cores and
              (b) for 1200 $\times$ 1200 $\times$ 1200 unit cells from 32 to 1728 cores.
              \label{Figure05}
		      }
	\end{figure}

  Finally, the efficiency of the kMC approach was evaluated 
  for vacancy diffusion in the SrTiO$_{2.75}$H$_{0.25}$ system. 
  For this purpose, the parallel efficiency of vacancy diffusion 
  was evaluated with 600$^3$~unit cells from a single core to 216 cores 
  (where the number of fragments ranged from 10 to 2160). 
  The results are shown in Figure~\ref{Figure05}(a).
  If the number of fragments and the number of CPUs are small,
  the calculation requires to update the respective event information
  for the entire system, and then a large amount of 
  atomic information has to be stored and accessed, operations that take additional 
  computational time. 
  Thus, the simulation with a small number of CPU cores 
  was indeed inefficient.
  Increasing the number of CPU cores 
  drastically reduced the computational time
  (Figure~\ref{Figure05}(a)). 
  In the simulation with more than 32 cores, 
  the computational time showed almost perfect parallel 
  efficiency.
  The computational time with 1 CPU core is 2850  min, 
  and the computational time is reduced to 3.1 min with 32 CPU cores,
  and 0.6944 min with 216 CPU cores.
  (140 \% parallel efficiency estimated the computational 
  time between 32 CPU and 216 CPU).

  For further evaluation of the feasibility of the proposed kMC approach, 
  the  parallel efficiency with a relatively large 
  system of 1200$^3$ unit cells was evaluated from 32 to 1800 cores 
(Figure~\ref{Figure05}(b)).
 Like with the parallel efficiency shown in Figure~\ref{Figure05}(a),
  the computational time could be reduced to 3.61 and 2.6 min 
  with 512 and 1728 CPU cores, and we achieved a reduction 
  in computational time by a factor of 24 and 33 in comparison with 32 CPU cores,
  and the parallel efficiency of 512 and  1728 CPU cores was 
  148.6  and  61.1\%. 
  Comparison of the parallel efficiency with the other simulation software 
  is also listed in TABLE~\ref{TABLEpara}. 
  The parallel efficiency in this study shows reasonable performance 
  in comparison with the other previous studies. 
  Thus, the developed fragment kMC approach could be successfully 
  applied to large-scale atomic diffusion
   by using a high-performance computing system. 
  Note that the parallel efficiency is not main factor for dertermining
  the quality of kMC soft, and another merit of fragment kMC 
  is the simple structure of the method 
  (The details will be shown in https://github.com/hiroyanakata/kMC.v01).

\subsection{Oxygen vacancy diffusion with dopant and defect interaction}

   It is also important to treat the interaction 
   between vacancy and dopant in kMC model, because the diffusion coefficients 
   strongly depend on the dopant.      
   For this purpose, the transition rate in Eq.~\ref{RateCnst1} is 
   reformulated based on the recent review of Koettgen et. al.\cite{ceria02}: 
   \begin{align}
     k_i^I = A 
         \mathrm{exp}
         \left[
           - \frac{E_a + \Delta E^{\mathrm{site}}/2}{RT} 
         \right],
   \end{align}
   where the $\Delta E^{\mathrm{site}}$ is potential energy difference between
   the final and initial vacancy site, 
   and the energy for each site $E^{\mathrm{site}}$ is calculated as follows:
   \begin{align}
    E^{\mathrm{site}}
  =& \sum_{i}^{n_{\mathrm{type}}} 
     \sum_{j}^{m_\mathrm{neighbor}}
     N_{i,j} E^{\mathrm{site}}_{i,j},
   \end{align}
     where $n_{\mathrm{type}}$ is number of different atom types  
     (e. g. dopant-vacancy, and vacancy-vacancy interactions),
     and $m_\mathrm{neighbor}$ is number of nearest neighbors sites
     taken into account in the simulation model,
     and $E^{\mathrm{site}}_{i,j}$ is the potential energy for respective 
     site positions, which are evaluated by first principle method.
      
   In this section, the effectiveness of the additional interaction potential 
   in the fragment kMC model is evaluated using Fe dopant in SrTiO$_3$\cite{STOFeexpt}
   as for the pilot test.
   The potential energy is estimated by first principle simulation 
   using CASTEP program\cite{CASTEP1,CASTEP2} with 
   Perdew-Burke-Ernzerhof (PBE+U) functional.\cite{PBE01,PBE02,HubbardU} 
   The Hubbard parameter U for Ti and Fe are set to U = 3 eV 
   and U = 5 eV, respectively.\cite{SrTiFeSim} 
   The results of V--Fe interaction and V--V interaction  
   are shown in TABLE~\ref{TABLEadd01}.

   Using the obtained potential energy interactions,
   the fragment kMC simulation is performed,
   and the results of diffusion coefficient are 
   compared between with and without the Fe dopant.
   The concentration of Fe and oxygen vacancies are 0.1 \%, 
   which is slightly higher concentration than actual experiment\cite{STOFeexpt}
   to evaluate the dopant effect of the diffusion coefficient clearly.

\begin{figure}
      \begin{center}
       \includegraphics[clip,width=12.0cm]{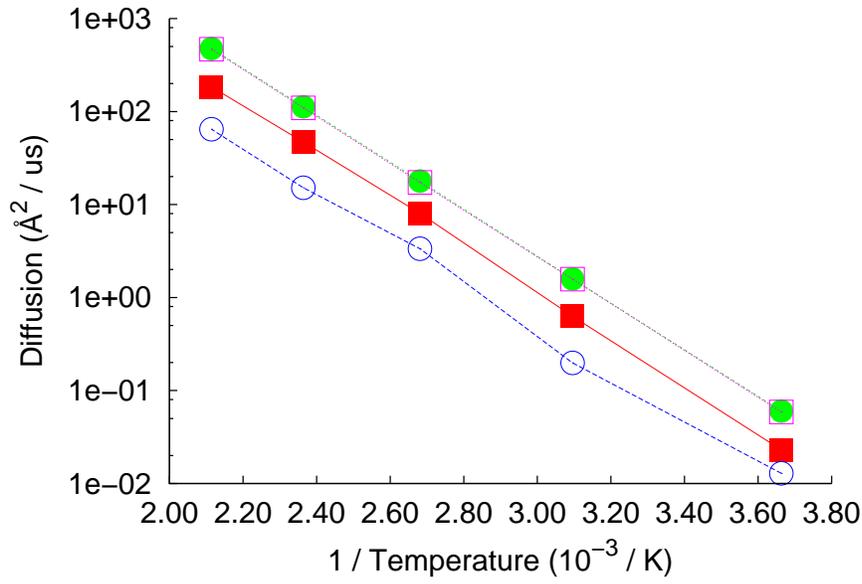} \\
      \end{center}
      \caption{
          Diffusion coefficient for doped SrTiO$_3$.
          Red closed square is diffusion coefficients 
          of Fe doped SrTiO$_3$ with Sr-V and V-V interaction,
          and blue open circle is  those  
          with Sr-V interaction only.
          Green closed circle is the diffusion coefficients 
          of pure SrTiO$_3$ with V-V interaction, 
          and magenta open square is those with no interaction.
          \label{Figureadd01}
	      }
\end{figure}

   The results of diffusion coefficients are shown in Figure~\ref{Figureadd01}.
   The oxygen vacancy diffusion coefficient in pure SrTiO$_3$ is 
   depicted in magenta open square in Figure~\ref{Figureadd01}. 
   Inclusion of V--Sr interaction drastically reduced the oxygen vacancy 
   diffusion coefficient (See blue open circle).
   It should be noted that the consideration of only V--Fe interaction 
   overestimated the reduction of vacancy diffusion. 
   Thus, additional inclusion of V--V interaction is significantly important 
   to estimate diffusion coefficients of Fe doped SrTiO$_3$.
   The comparison between red closed square  and blue open circle 
   shows that the inclusion of V--V interaction accelerate the vacancy 
   diffusion, which suggest the vacancy diffusion mechanism as follows:
   First, each Fe atom can trap oxygen vacancy.
   Then the trapped oxygen vacancies act as a repulsion force 
   to the other oxygen vacancies, which results in the acceleration 
   of oxygen vacancy diffusion. 
   Such interactions (V--Fe, and V--V) can be reasonably considered 
   in the fragment kMC method.

   We also investigate the effect of V--V interaction 
   on the diffusion coefficient in pure SrTiO$_3$.
   Then we do not observed much difference between with and without the V--V interaction
   in pure SrTiO$_3$ 
   (green closed circle and magenta open square in Figure~\ref{Figureadd01}). 
   Because the concentration of oxygen vacancy is low\cite{STOFeexpt}, 
   it is seldom to meet the oxygen vacancy with each other.

\subsection{Oxygen and hydride ions diffusion coefficients in SrTiO$_3$ and BaTiO$_3$}
   Another advantage of the fragment kMC programme is the efficient 
   evaluation of the diffusion coefficient for independent hydride or oxygen 
   ions. 
   Because the number of vacancies is significantly lower than the number of 
   oxygen and hydride ions, the standard kMC approach generates 
   an event list based on the vacancy, so the event selection 
   process can be significantly reduced in comparison with the explicit 
   update of oxygen or hydride ions. 
   By contrast, the fragment kMC randomly selects the possible event based on 
   the atoms in the system of interest, and therefore the event selection 
   scheme can be extended not only to vacancy diffusion, but also to 
   the explicit diffusion of oxygen or hydride ions.

   As for the pilot test of the fragment kMC approach, the oxygen and 
   hydride ion diffusion coefficients were evaluated for both SrTiO$_{(3-x)}$H$_x$ 
   and BaTiO$_{(3-x)}$H$_x$, where the hydride ions concentrations
   were 0.25, 0.35, and 0.45, respectively.
   To evaluate the apparent activation barrier for the respective hydride
   and oxygen ions, the simulations were performed for  
   550, 600, 650, and 700~K; thus, 24 types of kMC simulations were 
   performed to evaluate the respective activation barriers.
     
\begin{figure}
          \begin{center}
           \includegraphics[clip,width=12.0cm]{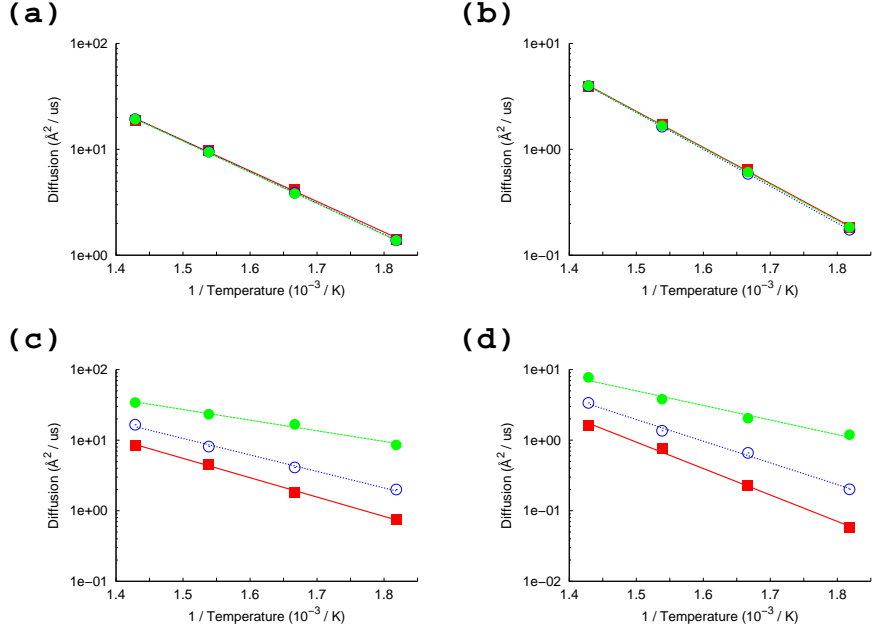} \\
          \end{center}
	      \caption{
              Oxygen and hydride ions diffusion coefficients from 550 to 700 K
              for (a) and (c)  SrTiO$_{(3-x)}$H$_x$ 
          and for (b) and (d)  BaTiO$_{(3-x)}$H$_x$.
              Red   closed squares, 
              blue  open circles, 
          and green closed circles illustrate the simulation results for $x=$ 0.25, 0.35, 
              and 0.45, respectively.
              The oxygen vacancy diffusion coefficients are shown in (a), and (b),
              while the hydrogen diffusion coefficients are shown in (c), and (d).
              \label{Figure06}
		      }
	\end{figure}

   The results for the oxygen diffusion coefficients are shown 
   in Figure~\ref{Figure06}(a) and (b) for SrTiO$_{(3-x)}$H$_x$ and 
   BaTiO$_{(3-x)}$H$_x$, respectively.
   In both cases, the oxygen diffusion is the rate determining step 
   for vacancy diffusion, and thus the diffusion coefficients 
   of oxygen vacancy do not vary with the concentration of hydride ions.

   The diffusion coefficient of hydride ions 
   was evaluated, and the results were compared 
   with the diffusion coefficient of oxygen. 
   The results for the hydride ions diffusion coefficient are shown 
   in Figure~\ref{Figure06}(c) and (d) for SrTiO$_{(3-x)}$H$_x$ 
   and BaTiO$_{(3-x)}$H$_x$, respectively. 
   As shown in Figure~\ref{Figure06}(c) and (d), 
   the diffusion coefficients vary significantly
   with hydride ions concentration ($x=0.25, 0.35, 0.45$). 

   The hydride ions diffusion is the vacancy-mediated diffusion,
   and this result can be understood based on the change of 
   the correlation factor\cite{liu2019highly}.
   The correlation factor in eq.~\ref{Cfactor} changes from $2.40\times10^{-3}$
   to $4.30\times10^{-3}$ with the increase of hydride ions concentration 
   from 0.25 to 0.45, which suggest that the vacancy mediated 
   hydrogen diffusion is significantly slowed down in the case of low
   concentration of hydrogen. The result estimated by kMC is consistent 
   compared to the analysis of the correlation factor approach.

   To evaluate the diffusion in more detail,
   the apparent activation barrier of hydride ions was evaluated for each 
   hydride ions concentration ($x=$ 0.25, 0.35, and 0.45).
   The results for the apparent activation barrier from the kMC simulation 
   are summarized in Table~\ref{TABLE02}.
   The activation barriers were 0.545 and 0.700~eV for SrTiO$_{2.75}$H$_{0.25}$ 
   and BaTiO$_{0.75}$H$_{0.25}$, respectively.
   Thus, the obtained apparent activation barriers also suggest 
   that oxygen migration blocks the diffusion of hydride ions 
   when the concentration of hydride ions is low ($x=$ 0.25).     
   Then, the activation barriers decrease with the increase in
   hydride ions concentration; the apparent activation barriers for $x=0.45$
   were 0.300 and 0.410~eV in SrTiO$_{(3-x)}$H$_x$ and BaTiO$_{(3-x)}$H$_x$, 
   respectively.

\subsection{Oxygen diffusion in $\mu$-scale BaTiO$_3$ model}
   To demonstrate the effectiveness of the fragment kMC approach, 
   the $\mu$-scale BaTiO$_3$ model is constructed,
   and the oxygen vacancy diffusion is evaluated. 
   The BaTiO$_3$  is a material that is important for industry, 
   and the oxygen vacancy diffusion under voltage application 
   has a crucial role in insulation deterioration.
   Thus, the oxygen vacancy diffusion in BaTiO$_3$
   was investigated to demonstrate the effectiveness of the fragment kMC approach.
   
   Likewise with the other ceramic materials,
   we practically use the polycrystalline BaTiO$_3$ for MLCC,
   whose grain boundary structures have not yet been 
   understood in detail.
   In the simulation perspective, it is also difficult 
   to make such a complex grain boundary structures 
   at current stage of the fragment kMC approach. 
   As for the initial pilot test whether the $\mu$-scale 
   oxygen vacancy diffusion can be evaluated or not,
   the simulation models are separated into respective grains,
   and we introduce an artificial intermediate BaTiO$_3$
   layer to separate the respective grains, and 
   the transition of oxygen vacancies inside the grain  are investigated. 
   The details of the test model are described as follows.

\begin{figure}
          \begin{center}
           \includegraphics[clip,width=6.0cm]{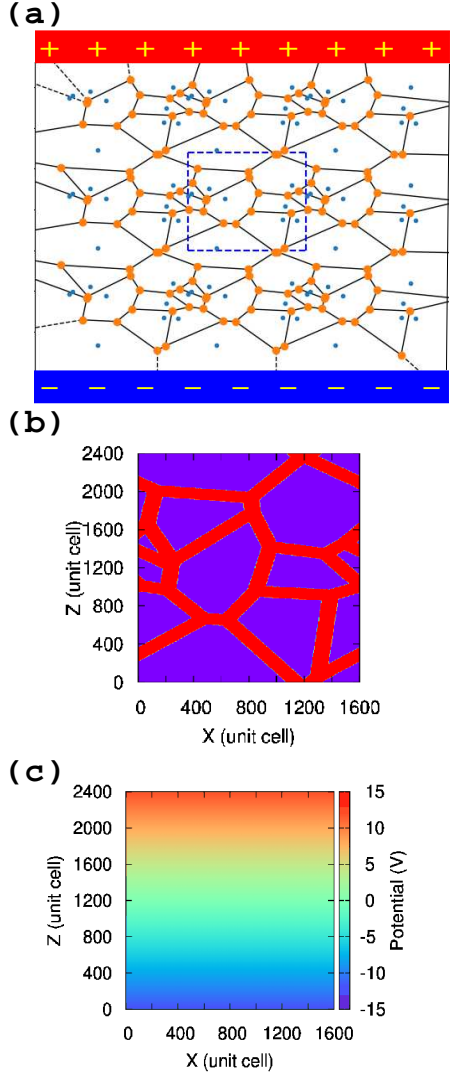} \\
          \end{center}
	      \caption{
              Simulation model for polycrystalline BaTiO$_3$.
              (a) Overview of the periodic simulation model used in this study, where
                  the blue dotted line denotes the actual simulated area.
              (b) Detailed simulation model for oxygen diffusion analysis,
                  where the purple colour denotes the respective grain, 
                  and the oxygen vacancy concentration is evaluated for each 
                  grain area. The red colour denotes the intermediate layer 
                  between the grains.
              (c) Schematic illustration of applying bias voltage
                  and  electrostatic potential.
              \label{Figure07}
		      }
	\end{figure}

   The Voronoi tessellation method\cite{Voronoi} was used
   to make the geometry of BaTiO$_3$  test model,
   and the simulation model of this study are shown 
   in Figure~\ref{Figure07}(a). 
   The test model is constructed in two dimensions (Figure~\ref{Figure07}) 
   for simplicity, but the actual kMC simulation was performed 
   in three dimensions.
   The white colour denotes 
   the grain, the black solid line is the grain boundary,
   and the orange circle is the edge of the grain boundary.
   The test model was constructed to satisfy 
   the periodic boundary condition (PBC), and the centre region
   depicted with an open rectangular (blue dotted line) was 
   explicitly treated by kMC, where the size of the simulation model 
   was 641.6~nm $\times$ 160.4~nm $\times$ 962.4~nm
   and the model contained 7,680,000,000 atoms in the system.
  
   In this study,
   the activation barrier over the respective boundary region was 
   set higher than those in the bulk area, 
   by adding positive electrostatic potential (0.2~eV). 
   The value 0.2~eV was determined from the 
   reported experimental study for oxygen 
   vacancy diffusion\cite{chang2018applicability},
   and thus, the total activation barrier through the boundary was 
   set to 0.9~eV.
   As noted,
   the additional intermediate layers are set nearby the grain 
   boundaries to remove the effect of vacancy transition over the grains
   as shown in Figure~\ref{Figure07}(b). 
   The analysis concentrated on the oxygen vacancy diffusion 
   within the respective grains,
   and we define the region between the grains and 
   the additional intermediate layer as the layer boundary region.

   In this study, the electrostatic potential was considered 
   by solving the Poisson equation\cite{PhaseField00}. 
   To apply the bias voltage, the boundary conditions were set 
   to 10.0~eV for the top of the $z$-axis 
   and $-$10.0~eV for the bottom of the $z$-axis, and the PBC was applied 
   for the $x$- and $y$-axes to solve the Poisson equation.
   The electrostatic potential after applying the bias voltage is 
   shown in Figure~\ref{Figure07}(c). 
   Because we applied the bias voltage along the $z$-axis, the PBC 
   of the $z$-axis is broken. Instead of the PBC, the open boundary 
   approximation was applied. If the oxygen vacancy diffuses over the boundary,
   we consider that the same amount of oxygen vacancies enters from 
   the other side of the boundary region. This approximation maintains 
   the total number of oxygen vacancies. 
    
\begin{figure}
          \begin{center}
           \includegraphics[clip,width=12.0cm]{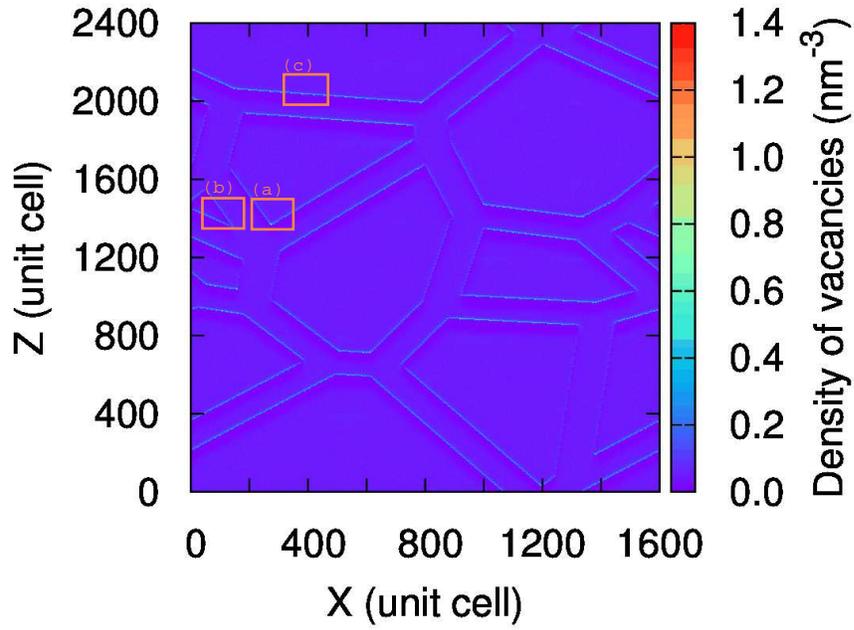} \\
          \end{center}
	      \caption{
             The oxygen vacancy concentration of the simulation model 
             in Figure~\ref{Figure07}(b). 
             Detailed analysis was performed for the orange open square areas 
             (a), (b), and (c). See the main text and Figure~\ref{Figure09} 
             for detail. 
             \label{Figure08}
		  }
	\end{figure}

   The distribution of oxygen vacancies obtained by a 5~ms simulation
   is shown in Figure~\ref{Figure08}. 
   It can be observed that the oxygen vacancies tend to gather around
   the bottom of the layer boundary (The charge of vacancy is 
   compensated by the reduction of Titanium from IV to III). 
   The highest oxygen vacancy concentration can be found  
   around $x=138$ and $z=686$, 
   and the density of oxygen vacancy in the highest region is 1.4~nm$^{-3}$.
   The average vacancy concentration in this study was 0.0465~nm$^{-3}$,
   and an oxygen vacancy concentration about 30 times larger
   was observed.
   It should be noted that we have also found some oxygen 
   vacancy concentration in the bottom of intermediate layers 
   (nearby the top of grain),
   because of the artificial separation of respective grains.
   More detail information for the oxygen vacancy concentration 
   in the intermediate layers are shown  in supplementary materials,
   and here we focus on the oxygen vacancy concentration inside the grain.
       
\begin{figure}
          \begin{center}
           \includegraphics[width=8.0cm]{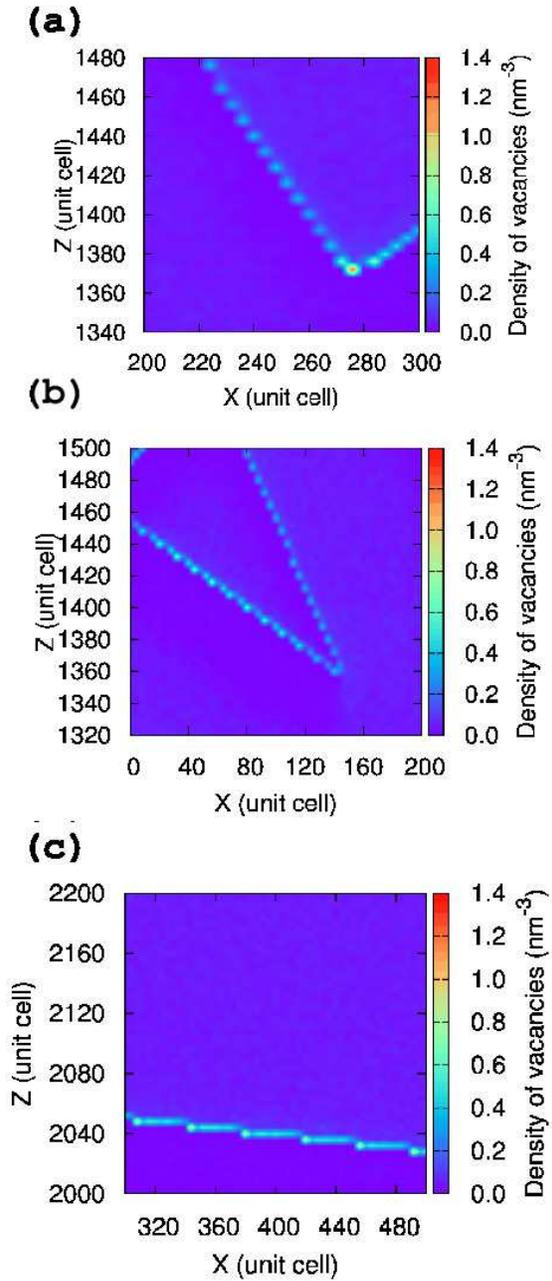} \\
          \end{center}
	      \caption{
             Oxygen vacancy concentration of the respective areas shown in
             Figure~\ref{Figure08}(a), (b), and (c). 
             \label{Figure09}
		  }
	\end{figure}

   The highest oxygen vacancy concentration can be found 
   near the triangular layer boundary on the bottom of Figure~\ref{Figure08}(a), 
   and the detailed distribution of oxygen vacancies is shown in 
   Figure~\ref{Figure09}(a).  As shown in Figure~\ref{Figure09}(a), 
   the oxygen vacancies diffuse to increasingly low regions,
   but the layer boundary area prohibits to transfer the vacancy further,  
   and the vacancy migrates to the next lower layer boundary area. 
   As a result, the oxygen 
   vacancies gather to the lowest edge area (See Figure~\ref{Figure09}(a)).

   The grain size is also an important factor to determine 
   the oxygen vacancy density  in the edge area. 
   Figure~\ref{Figure09}(b) and Figure~\ref{Figure08}(b)  show 
   the oxygen vacancy density at the edge area in the case of 
   small grain size.
   The total number of oxygen vacancies 
   in small grains is lower than in large grains,
   which also results in a smaller vacancy concentration 
   in the edge area.
   Compared with the results in Figure~\ref{Figure09}(a),
   the edge in a small grain (Figure~\ref{Figure09}(b))
   has a lower oxygen vacancy concentration, 
   and the concentration of oxygen vacancies is about 35.7\%  
   lower than the highest oxygen vacancy concentration.

   Another type of oxygen vacancy concentration can be found 
   in the region where the layer boundary is a horizontal 
   line (parallel to the $x$-axis, see Figure~\ref{Figure08}(c)). 
   The distribution of oxygen vacancies around the horizontal layer boundary 
   is shown in  Figure~\ref{Figure09}(c), where it can be observed that
   the number of oxygen vacancies is lower than that of the edge area,
   even for the largest grain size (Figure~\ref{Figure08}(c)). 

   In summary, the simulation results indicate 
   that the oxygen vacancy diffusion processes occur as follows. 
   The oxygen vacancies tend to be trapped in the layer boundary area, and 
   going through the boundary region, the vacancies are finally trapped
   at the edge area. Thus, without the edge area 
   (i.e., the layer boundary is parallel to the $x$-axis), 
   the oxygen vacancy concentration is very low.
   The oxygen vacancy migration inside the respective grains 
   is considered to be reasonable based on the general 
   behaviour of oxygen vacancy migration in grains
   \cite{strukov2012thermophoresis,yang2004oxygen,chazono2001dc}, 
   and  the above pilot test suggest that  
   the developed fragment kMC can simulate 
   the oxygen migration in such a large $\mu$ scale BaTiO$_3$ model 
   with a fully atomistic computational model.

\section{Conclusion}
    In this study, the fragment kMC method is proposed.
    This approach introduces the auxiliary value $R^\mathrm{max}$, 
    and the event selection can be decomposed into atomic-based fragments.
    The kMC method can be easily parallelized 
    towards calculation of atomic migrations. 
    The diffusion coefficients perfectly agree with the conventional 
    kMC simulation, and the parallel efficiency of fragment kMC was 
    good enough until the number of CPU cores reached 1728.

    The method can be straightforwardly extended to analyse 
    the diffusion coefficient for independent atomic types.
    The respective hydride ions and oxygen diffusion coefficients
    were evaluated in perovskite  SrTiO$_{(3-x)}$H$_x$ and BaTiO$_{(3-x)}$H$_x$ systems.
    The simulation was performed with various hydride ions
    compositions ($x=$ 0.25, 0.35, 0.45),
    which were slightly difficult to predict using conventional kMC.
    Then, the simulation  provided good prediction of the 
    diffusion coefficient of hydride ions at reasonable computational cost.

    Finally, the fragment kMC was extended to simulate the $\mu$ scale 
    BaTiO$_3$ under bias voltage application. 
    We found that the oxygen vacancy concentration in the edge region
    was the highest, and the obtained results were reasonable 
    based on the general behaviour of oxygen migration in  
    BaTiO$_3$ materials.
    Thus, we conclude that the developed fragment kMC programme can be useful 
    for simulating many kinds of atomistic diffusion in crystal systems,
    and we hope that the proposed method can be widely used to understand 
    the diffusion mechanism in the field of materials and chemistry.

\section*{ACKNOWLEDGMENT}
We thank 
R.I.I.T.\ at Kyushu University (Japan) for providing computational resources. 
This research also used computational resources of 
the Fujitsu PRIMERGY CX400M1/CX2550M5(Oakbridge-CX) 
by Information Technology Center, The University of Tokyo
through the HPCI System Research project (Project ID:hp200015).

\bibliographystyle{aip}
\bibliography{FkMC}
   
\newpage

TABLE captions.


\begin{table}[h!]
\caption[]{
   Average and standard deviation (SD) of the oxygen vacancy diffusion 
   coefficient (\AA$\mu$s) in SrTiO$_3$ single crystal.
\label{TABLE01}}
\begin{tabular}{lrr}\hline
   number of fragment &   Average         &   (SD)      \\\hline
     1                &     6433          &   56        \\
     8                &     6428          &   75        \\
    32                &     6433          &   54        \\     
    64                &     6385          &   59        \\     
   125                &     6412          &   49        \\     
   216                &     6391          &   66        \\\hline
\end{tabular}
\\
\end{table}%

\begin{table}[h!]
\caption[]{
    Parallele efficiency comparison with previous studies.
\label{TABLEpara}}
\begin{tabular}{lrr}\hline
   Software                  &   CPU cores       &  efficiency  \\\hline
  LAKIMOCA\cite{LAKIMOCA}    &  Only Serial      &   N. A.      \\
  mesokMC\cite{tada2013}     &       32          &   55         \\
  Crystal-KMC\cite{cKMC}     &      800          &   56         \\     
  spKMC\cite{spKMC}          &      256          &   82         \\     
  this study                 &      512          &  148         \\\hline
\end{tabular}
\\
\end{table}%

\begin{table}[h!]
\caption[]{
  The potential energy between Vacancy and Fe (left column),
  and potential energy between Vacancy and Vacancy (right column)
  estimated by first principle simulation.
  \label{TABLEadd01}
}
\begin{tabular}{lrrrr}\hline
site  &   Dis    &     V-Fe  &       Dis  &    V-V\\\hline   
 1st  & 1.952    &   0.000   &     2.76   &  0.343\\\
 2nd  & 4.390    &   0.510   &     3.91   &  0.235\\\
 3rd  & 5.920    &   0.687   &     4.78   &  0.200\\\
 4th  & 7.030    &   0.709   &     5.52   &  0.180\\\
 5th  &  N.A.    &   N.A.    &     6.17   &  0.023\\\
 6th  &  N.A.    &   N.A.    &     7.81   &  0.000\\\hline
\end{tabular}
\end{table}%

\begin{table}[h!]
\caption[]{
    Apparent activation barrier (eV) estimated by the diffusion 
    coefficient of hydride ions.
    The simulation results are shown in Figure~\ref{Figure06}.
    \label{TABLE02}
}
\begin{tabular}{lrr}\hline
   H concentration &   SrTiO$_3$        &   BaTiO$_3$      \\\hline
     0.45          &   0.300            &  0.410           \\
     0.35          &   0.464            &  0.611           \\
     0.25          &   0.545            &  0.700           \\\hline
\end{tabular}
\end{table}%

\newpage
\section*{Figure captions}
\newpage

%

\newpage

\newpage

\newpage

\newpage

\newpage

\newpage

\newpage

\newpage

\newpage

\newpage

\end{document}